\documentstyle[eqsecnum,aps]{revtex}

\input{epsf.tex}
\begin{document}
\draft
\twocolumn[\hsize\textwidth\columnwidth\hsize\csname
@twocolumnfalse\endcsname
\renewcommand{\thefootnote}{\roman{footnote}} 
\title{Diffusion and scaling in escapes from two-degree-of-freedom Hamiltonian 
systems}
\author{Henry E. Kandrup\footnote{Electronic mail: kandrup@astro.ufl.edu}}
\address{ Departent of Astronomy, Department of Physics, and
Institute for Fundamental Theory \\
University of Florida, Gainesville, Florida 32611}
\author{Christos Siopis\footnote{Electronic mail: siopis@astro.ufl.edu}}
\address{Department of Astronomy, University of Florida, Gainesville, Florida 
32611 \\
Institut f\"ur Astronomie, Universit\"at Wien, T\"urkenschantzstra{\ss}e 17,
A-1180, Wien, AUSTRIA}
\author{G. Contopoulos\footnote{Electronic mail: 
gcontop@atlas.uoa.ariadne-t.gr}}
\address{Astronomy Department, University of Athens, 
Panepistimiopolis, 157 83,  Athens, GREECE}
\author{Rudolf Dvorak\footnote{Electronic mail: 
dvorak@astro.ast.univie.ac.at}}
\address{Institut f\"ur Astronomie, Universit\"at Wien, 
T\"urkenschantzstra{\ss}e 17, A-1180, Wien, AUSTRIA}
\date{\today}
\maketitle
\begin{abstract}
This paper summarises an investigation of the statistical properties of orbits 
escaping from three different two-degree-of-freedom Hamiltonian systems which
exhibit global stochasticity. Each 
time-independent $H=H_{0}+{\epsilon}H'$, with $H_{0}$ an integrable 
Hamiltonian 
and ${\epsilon}H'$ a nonintegrable correction, not necessarily small. 
Despite possessing very different symmetries, ensembles of orbits in all three 
potentials exhibit similar behaviour. For ${\epsilon}$ below a critical 
${\epsilon}_{0}$, escapes are impossible energetically. For somewhat higher 
values, escape is allowed energetically but still many orbits never escape. 
The escape probability $P$ computed for an arbitrary orbit ensemble decays 
towards zero exponentially. At or near a critical value 
${\epsilon}_{1}>{\epsilon}_{0}$ there is a rather abrupt qualitative 
change in behaviour. Above ${\epsilon}_{1}$, $P$ typically exhibits (1) an 
initial rapid evolution towards a nonzero $P_{0}({\epsilon})$, 
the value of which is independent of the detailed choice of initial conditions,
followed by (2) a much slower subsequent decay  towards zero which, in at 
least one case, is well fit by a power law $P(t){\;}{\propto}{\;}t^{-\mu}$, 
with ${\mu}{\;}{\approx}{\;}0.35-0.40$. In all three cases, $P_{0}$ and the 
time $T$ required to converge towards $P_{0}$ scale as powers of 
${\epsilon}-{\epsilon}_{1}$, i.e., 
$P_{0}{\;}{\propto}{\;}({\epsilon}-{\epsilon}_{1})^{\alpha}$
and
$T{\;}{\propto}{\;}({\epsilon}-{\epsilon}_{1})^{\beta}$, 
and $T$ also scales in the linear size $r$ of the region sampled for
initial conditions, i.e., 
$T{\;}{\propto}{\;}r^{-\delta}$. To within statistical uncertainties, the best 
fit values of the critical exponents ${\alpha}$, ${\beta}$, and ${\delta}$ 
appear to be the same for all three potentials, namely 
${\alpha}{\;}{\approx}{\;}0.5$, ${\beta}{\;}{\approx}{\;}0.4$, and
${\delta}{\;}{\approx}{\;}0.1$, and satisfy
${\alpha}-{\beta}-{\delta}{\;}{\approx}{\;}0$. 
The transitional behaviour observed near ${\epsilon}_{1}$ is attributed to the
breakdown of some especially significant {\it KAM} tori or cantori. The power 
law behaviour at late times is interpreted as reflecting intrinsic diffusion 
of chaotic orbits through cantori surrounding islands of regular orbits.
\end{abstract}
]
\narrowtext
\section{INTRODUCTION}
\label{sec:level1}
This paper summarises a study of the problem of escapes of energetically 
unbound orbits in strongly non-integrable two-degree-of-freedom Hamiltonian
systems as an example of phase space transport in complex systems.
This work has led to two significant conclusions: (1) When evolved into the 
future,
ensembles of orbits of fixed energy often exhibit a rapid approach towards a 
constant escape probability $P_{0}$, the value of which is independent of 
the details of the ensemble and exhibits interesting 
scaling behaviour. 
Moreover, the values of the critical exponents appear to be relatively 
insensitive to the choice of Hamiltonian. (2) At later times, the escape 
probability decreases in a fashion which, for at least one model system, is 
well fit by a power law $P(t){\;}{\propto}{\;}t^{-\mu}$ with 
${\mu}{\;}{\approx}{\;}0.35-0.40$. This nontrivial time-dependence is 
attributed to the fact that the possibility of escape to infinity is 
controlled by cantori, which can trap chaotic orbits near regular 
regions for extremely long times.

The first three papers in this series\cite{1,2,3} (hereafter 
Papers $1$ - $3$) described a numerical investigation 
of the statistical properties of orbit ensembles evolving in nonintegrable 
two-degree-of-freedom Hamiltonian systems where it is possible energetically 
for trajectories to escape to infinity. Earlier investigations of individual
orbits in these systems had led to two significant conclusions\cite{4,5}:
(1)
Just because escape is possible energetically does not mean that escape 
will inevitably occur; and, even if escape does occur, the time required for a
trajectory to cross a Lyapunov curve\cite{Church} and hence escape to 
infinity may be very long compared with the natural crossing time. (2) In some
phase space regions, 
the time and direction of escape vary smoothly as a function of initial 
conditions, but in other cases one discovers instead an apparent near-fractal
dependence on the specific choice of initial data (cf.\cite{Henon}). 

This problem of escapes is an example of phase space transport in complex
two-degree-of-freedom Hamiltonian systems, a subject which has been explored
in detail over the past two decades (cf. \cite{Wiggins,Meiss,Reichl,LL} and
references cited therein). The passage of orbits through Lyapunov curves and 
their subsequent escape to infinity is the most conspicuous aspect of the
transport, but crucial features of the bulk flow, especially at late times,
appear to be controlled by diffusion through 
cantori\cite{Percival,Aubry,Mather,Shirts}, which can trap orbits for very
long times (cf. \cite{Meiss}). The chaotic behaviour
of late escapers indicates that this problem is closely related to chaotic 
scattering (cf. \cite{Ott1,Ott2}), where an incident trajectory scatters to 
infinity 
at a time and in a direction that can exhibit a fractal dependence on the
value of the impact parameter. However, the problem of escapes is also related
to a variety of other physical problems, including, e.g., the dissociation of
molecules (cf. \cite{Noid}) or the evaporation of stars from a cluster (cf.
\cite{Heggie}). 
Indeed, the phase space interpretation of the escape problem suggested in
Section IV 
is completely consistent with the detailed phase 
space description deduced recently for escapes in the planar isoceles 
three-body problem.\cite{Zare}

 
Paper I showed that, for at least one particular Hamiltonian system (the
$H_{1}$ of eq.~[1]), the microscopic chaos exhibited by individual orbits 
leads macroscopically to bulk regularities: (1) For sufficiently large
deviations from integrability, localised ensembles of initial conditions 
evolve so as to exhibit a rapid approach towards a near-constant escape 
probability $P_{0}$, which is independent of the specific choice of initial 
conditions and which, if at all, only changes on a significantly longer time 
scale. (2) The value of $P_{0}$ scales in terms of an ``order parameter'' 
${\epsilon}-{\epsilon}_{1}$. (3) For ensembles that probe a phase space region 
of specified size $r$, the time $T$ required to converge towards $P_{0}$ also 
scales in ${\epsilon}-{\epsilon}_{1}$. (4) For fixed values of ${\epsilon}-
{\epsilon}_{1}$, $T$ also scales in the size $r$ of 
the phase space region sampled by the initial ensemble.

Subsequent work, summarised and extended here, has addressed several 
questions not considered in Paper 1:
\par\noindent
(1) Do other potentials exhibit similar behaviour and, if so, are the scaling
exponents the same? In other words, {\it could this behaviour be universal}?
\par\noindent
(2) What happens at much later times? Does $P$ remain constant or is there
a different asymptotic behaviour for larger values of $t$?
\par\noindent
(3) Is there any obvious correlation between the time of escape for individual
orbits in the ensemble and the exponential instability of those orbits, as 
probed, e.g., by short time Lyapunov exponents 
(cf.\cite{Grass,Badii,KAA,Voglis})?
\par\noindent
(4) Can one identify a simple, physically well motivated model to 
explain the observed behaviour?

This work is based on a detailed investigation of orbits in three different
Hamiltonians, namely:
$$H_{1}={1\over 2}({\dot x}^{2}+{\dot y}^{2}+x^{2}+y^{2})-{\epsilon}x^{2}y^{2},
\eqno(1) $$
$$H_{2}={1\over 2}({\dot x}^{2}+{\dot y}^{2}+x^{2}+y^{2})-{\epsilon}xy^{2},
\eqno(2) $$
and
$$H_{3}={1\over 2}{\Biggl(}{\dot x}^{2}+{\dot y}^{2}+x^{2}+y^{2}-
{2\over 3}y^{3}{\Biggr)}+{\epsilon}x^{2}y,  \eqno(3) $$
where ${\dot x}$ and ${\dot y}$ denote canonical momenta.
In each case, the Hamiltonian is of the form 
$$H=H_{0}+{\epsilon}H' ,\eqno(4) $$
with $H_{0}$ integrable and ${\epsilon}H'$ a nonintegrable correction. 
${\epsilon}$ is 
taken to be non-negative but is {\it not} assumed to be small. 
These Hamiltonians exhibit very different symmetries. $H_{1}$ is invariant
under $x\to -x$ and/or $y\to -y$ and has four identical escape channels. 
$H_{2}$ is only symmetric with respect to $y\to -y$, and has two channels of 
escape. For ${\epsilon}=1$, $H_{3}$ reduces to motion in the H\'enon-Heiles 
potential, which manifests a $2{\pi}/3$ rotation symmetry, but for 
${\epsilon}>1$ this discrete symmetry is broken.
Representative equipotential surfaces 
are exhibited in Fig. 1.

This research has led to four principal conclusions:
\par\noindent
(1) For all three systems, there exists a critical ${\epsilon}_{1}$,
larger than ${\epsilon}_{0}$, the smallest ${\epsilon}$ for which escapes
can occur, which signals a qualitative change in short time behaviour:
Below ${\epsilon}_{1}$ the escape probability $P$ decays towards zero 
exponentially, but for larger values of ${\epsilon}$  one observes 
instead an initial approach towards a near-constant nonzero $P_{0}$,
the value of which is independent of the detailed choice of initial conditions.
\par\noindent
2) For all three systems, $P_{0}$ scales in ${\epsilon}-{\epsilon}_{1}$, i.e., 
$P_{0}{\;}{\propto}{\;}({\epsilon}-{\epsilon}_{1})^{\alpha}$ with 
${\alpha}>0$. For a uniform sampling of a given phase space region of
fixed size $r$, the time $T$ required to converge towards $P_{0}$ also scales, 
i.e., $T{\;}{\propto}{\;}({\epsilon}-{\epsilon}_{1})^{-\beta}$, with 
${\beta}>0$. For fixed ${\epsilon}$, $T$ also scales in the linear size $r$ of 
the phase space region initially sampled, i.e.,
$T{\;}{\propto}{\;}r^{-\delta}$ with ${\delta}>0$. Finally, the data are 
at least consistent with the possibility  
that the numerical values of ${\alpha}$, 
${\beta}$, and ${\delta}$ are the same for all three potentials; and that 
${\alpha}-{\beta}-{\delta}=0$. 
\par\noindent
(3) At later times, $P$ deviates from $P_{0}$ by exhibiting a slow 
decrease towards zero. For at least one Hamiltonian, namely $H_{3}$, this 
later time evolution is well fit by a power law $P{\;}{\propto}{\;}t^{-\mu}$, 
with a positive constant ${\mu}<1$.
\par\noindent
(4) At least for $H_{3}$, and possibly for $H_{1}$ and $H_{2}$, 
orbits that escape early on tend to be more unstable than orbits which only 
escape at much later times, in that they have a larger short time
Lyapunov exponent. Computed distributions of short time Lyapunov
exponents and surfaces of section both support the hypothesis that the chaotic 
orbits divide, at least approximately, into relatively distinct subclasses,
presumably separated by cantori.
\par\noindent
The last two points suggest strongly that the late time evolution of $P(t)$ is 
controlled by cantori, which can trap chaotic orbits near regular islands for 
very long times. It is well known (cf. \cite{Meiss}) that cantori typically 
constitute the principal impediment for efficient phase space transport in 
two-degree-of-freedom systems and that diffusion through cantori is usually 
not characterised by a constant escape probability.

Section II describes the observed short time behaviour, summarising results 
presented in Papers 1 - 3 and discussing the evidence for universality. 
Section III focuses on longer time evolution, using surfaces of section and 
short time Lyapunov exponents to provide insights into flows associated 
with initially localised orbit ensembles. Section IV suggests a tentative 
physical interpretation of the numerical results in terms of flows in a 
chaotic phase space partitioned by cantori. Phase space is 
presumed to be dominated by ``unconfined'' chaotic orbits which, in the 
absence of any cantori or Lyapunov curves, would evolve towards a statistical 
equilibrium  (cf. \cite{LL}). The constant $P_{0}$ observed at relatively 
early times is attributed to the fact that orbits sampling this 
near-equilibrium will escape at a near-constant rate. The decaying $P(t)$ 
later on reflects the fact that the phase space also includes an appreciable 
measure of temporarily ``confined'' or ``sticky'' (cf. \cite{Shirts,Contop}) 
chaotic 
orbits which, albeit not trapped within the system forever, can only escape 
much later once they have breached one or more cantori to become unconfined.

\section{SHORT TIME BEHAVIOUR} 
\label{sec:level1}
\subsection{Description of the experiments}
\label{sec:level2}
The experiments described here entailed a study of orbit ensembles with fixed 
energy $h$ evolving in $H_{1}$, $H_{2}$, and $H_{3}$ with variable 
${\epsilon}$. Attention focused exclusively on ${\epsilon}>{\epsilon}_{0}(h)$, 
the smallest value of ${\epsilon}$ for which escape to infinity is possible 
energetically. The values of $h$ and ${\epsilon}_{0}$ for all three cases are 
given in Table 1.

Ensembles of initial conditions were generated by uniformly sampling a square 
cell of linear dimension $r$ in the $(x,{\dot x})$ plane, setting $y=0$, and 
then computing ${\dot y}={\dot y}(x,{\dot x},h)>0$. The cells were chosen to 
sample ``interesting'' phase space regions where most of the orbits do {\it 
not} escape at very early times. This implied that the time of escape for any 
given orbit typically manifested a sensitive dependence on 
initial conditions. Most of the computations involved a fiducial cell size 
$r=0.05$. However, when exploring the effects of varying cell size, cells as 
small as $r=1.0\times 10^{-4}$ were also used. 

Each initial condition was integrated into the future and the location of the 
orbit on a Poincar\'e section noted at successive consequents, i.e., 
successive crossings of the $x=0$ phase space hyperplane with ${\dot x}<0$. 
If after consequent $t-1$ but before consequent $t$ the orbit crossed one of 
the Lyapunov curves, the orbit was recorded as having escaped at $t$. The 
experiments with $H_{1}$ \cite{1,4,5} 
used a time series integrator. However, the experiments involving $H_{2}$ and 
$H_{3}$ \cite{2,3} exploited a more efficient Lie integrator truncated at 
twelfth order\cite{Dvorak}, which facilitated integrations of significantly 
larger 
ensembles typically containing $2000\times 2000$ initial conditions or more.

The fundamental object of interest is $P({\epsilon},t)$, the probability that
a randomly chosen orbit is an escaping orbit with escape occuring between 
consequents 
$t-1$ and $t$. This escape probability, along with an estimated 
uncertainty ${\Delta}P({\epsilon},t)$, was defined by the obvious relation
$$P({\epsilon},t){\pm}{\Delta}P({\epsilon},t)=
{N_{esc}{\pm}\sqrt{N_{esc}}\over N_{tot}} ,\eqno(5)$$
where $N_{esc}$ denotes the number of trajectories that escape between $t-1$ 
and $t$ and $N_{tot}$ the total number present at $t-1$.

\subsection{Results from the experiments}
\label{sec:level2}
At early times, the escape probability can exhibit a complex, highly irregular 
behaviour, the details of which depend sensitively on the size and location of 
the initial cell. However, at somewhat later times, $P$ tends instead to 
exhibit a more systematic behaviour that is seemingly independent of the 
initial cell and depends only on the value of ${\epsilon}$. The qualitative 
form of this behaviour depends crucially on whether ${\epsilon}$ is above or 
below a critical ${\epsilon}_{1}>{\epsilon}_{0}$. 

Below ${\epsilon}_{1}$, the escape probability $P({\epsilon},t)$ decays 
towards zero in a fashion that is well fit by an exponential. This is, 
e.g., illustrated in Figs. 2 a and b, which exhibit $P(t)$ and 
${\rm ln}\,P(t)$ for one value of ${\epsilon}<{\epsilon}_{1}$ in $H_{2}$, 
namely ${\epsilon}=1.04$. The solid curve superimposes a best fit exponential 
$P{\;}{\propto}{\;}\exp (-t/{\tau})$, with ${\tau}=62.0$.

For ${\epsilon}>{\epsilon}_{1}$, $P({\epsilon},t)$ appears instead to 
converge towards a nonzero $P_{0}({\epsilon})$, the value of which depends on 
${\epsilon}$ but is independent of 
the cell of initial 
conditions. The evidence is particularly compelling for $H_{1}$ and $H_{3}$, 
but somewhat weaker for $H_{2}$ where, especially for small 
${\epsilon}-{\epsilon}_{1}$, the convergence is relatively slow and can merge 
into the later time evolution described in Section III. Examples of this 
behaviour are provided in Figs. 3 a-c, which exhibit $P(t)$ for representative 
values ${\epsilon}> {\epsilon}_{1}$ in $H_{1}$, $H_{2}$, and $H_{3}$. (Other 
examples are provided in Papers 1 - 3.) The transition at or near
${\epsilon}_{1}$ is quite abrupt and, for ${\epsilon}>{\epsilon}_{1}$, 
$P_{0}({\epsilon})$ is a monotonically increasing function
of ${\epsilon}$. Moreover, for all three Hamiltonians one finds that, at least 
for relatively small values of the order parameter ${\epsilon}-{\epsilon}_{1}$,
the escape probability $P_{0}({\epsilon})$ exhibits a simple scaling, namely 
\cite{1,2,3}
$$P_{0}({\epsilon}){\;}{\propto}{\;}({\epsilon}-{\epsilon}_{1})^{\alpha},
\eqno(6) $$
with a constant ${\alpha}>0$. Illustrations of the goodness of fit of this 
scaling relation for the Hamiltonians $H_{1}$ and $H_{3}$ are provided, 
respectively, by Figs. 7 in Paper 1 and Figs. 4 in Paper 3, which 
exhibit plots of $P$ vs.~${\epsilon}$ and 
${\rm ln}\,P$ vs.~${\rm ln}\,({\epsilon}-{\epsilon}_{1})$. The values of 
${\epsilon}_{1}$ are again given in Table 1. 
That the escape probability approaches a roughly time-independent value can 
be interpreted by supposing that the cell of initial conditions has dispersed 
to fill certain large regions inside the Lyapunov curves in a nearly uniform 
fashion, and that trajectories are escaping near-randomly at a constant rate.

Using operational prescriptions described in Paper 1, one can also estimate
the time $T$ required for an ensemble with ${\epsilon}>{\epsilon}_{1}$ to
approach $P_{0}$. For cells of fixed size $r$, this $T$ appears to be roughly 
independent of initial conditions, depending only on ${\epsilon}$. Moreover, 
one finds that, for all three Hamiltonians, $T$ also scales in ${\epsilon}$ 
\cite{1,2,3}, i.e.,
$$T({\epsilon}){\;}{\propto}{\;}({\epsilon}-{\epsilon}_{1})^{-{\beta}},
\eqno(7)$$
with ${\beta}>0$. Illustrations of the goodness of fit of this 
scaling relation for the Hamiltonians $H_{1}$ and $H_{3}$ are provided, 
respectively, by Figs.~8 in Paper 1 and Figs.~5 in Paper 3, which 
exhibit plots of $T$ vs.~${\epsilon}$ and 
${\rm ln}\,T$ vs.~${\rm ln}\,({\epsilon}-{\epsilon}_{1})$. 

To the extent that a constant $P_{0}$ reflects a population that has dispersed 
throughout the regions inside the Lyapunov curves, one might anticipate that 
the convergence time $T$ would depend on
the size $r$ of the initial cell, smaller cells approaching 
$P_{0}({\epsilon})$ more slowly. This too was confirmed numerically, Indeed, 
for all three Hamiltonians one finds that, for a fixed value of ${\epsilon}$,
the convergence time $T(r)$ also scales in $r$ \cite{1,2,3}, i.e.,
$$T(r){\;}{\propto}{\;}r^{-\delta}, \eqno(8)$$
with ${\delta}>0$. This is illustrated in Figs.~9 and 10 of Paper 1
for two different values of ${\epsilon}$ for the Hamiltonian $H_{1}$.

It is obviously important to determine how abrupt the transition at
${\epsilon}{\;}{\approx}{\;}{\epsilon}_{1}$ actually is. However, this is
difficult numerically. Eq. (7) implies that $T({\epsilon})$ diverges for
${\epsilon}\to{\epsilon}_{1}$, but this critical slowing down implies an 
intrinsic limitation in one's ability to probe the details near the transition 
point. 

\subsection{Possible evidence for universality}
\label{sec:level2}
That the scaling relations (6) - (8) hold for all three Hamiltonians
is clearly interesting. Even more striking, however, is the fact that, as is
evident from Table 1, the values of the exponents ${\alpha}$, ${\beta}$, and 
${\delta}$ are very similar for all three systems. In each case,
$${\alpha}{\;}{\sim}{\;}0.5, \hskip .2in
{\beta}{\;}{\sim}{\;}0.4, \hskip .2in {\rm and} \hskip .2in
{\delta}{\;}{\sim}{\;}0.1. \eqno(9) $$

The uncertainties in ${\alpha}$ are dominated by uncertainties in the correct
value of ${\epsilon}_{1}$. As discussed in Paper 1, because of the 
aforementioned critical slowing down ${\epsilon}_{1}$ is best estimated by 
looking at somewhat higher values of ${\epsilon}$ and extrapolating to smaller 
${\epsilon}$. The uncertainties in ${\beta}$ are dominated by the 
precise prescription used to identify a convergence time. In particular, even 
though it is usually easy to determine a lower bound on the convergence time, 
the determination of an upper bound proves more difficult. It follows that the 
quoted error bars in Table 1 can be asymmetric. As regards the best fit 
${\delta}$, there are two principal sources of uncertainty, namely (1) that 
the effect is relatively small (so that the fractional error is large) and (2) 
that, especially for very small values of $r$, different ensembles can exhibit 
significant variability.

Given these uncertainties, one cannot conclude unambiguously that 
${\alpha}$, ${\beta}$, and ${\delta}$ are strictly equal
for all three Hamiltonians. However, one {\it can} conclude that
they are all comparable in size for all three systems and that, consistent
with the uncertainties, they may in fact be equal.

It is also true that, to within statistical uncertainties, 
$${\alpha}-{\beta}-{\delta}{\;}{\approx}{\;}0. \eqno(10)$$
For $H_{1}$ and $H_{3}$ the evidence for this assertion is relatively strong.
For $H_{2}$ the case is somewhat weaker, largely because the best fit 
${\beta}$ is somewhat larger than for the other two systems. However, it 
should be noted that the error bars on the ${\beta}$ for $H_{2}$ 
are especially big. This reflects the fact that, for this system, the short 
time behaviour described in this Section merges relatively quickly into the
later time evolution described in Section III, where $P(t)$ begins to decay to
values below $P_{0}$. For $H_{1}$ and $H_{3}$ this subsequent decay only 
become significant at somewhat later times, at least for larger values of
${\epsilon}-{\epsilon}_{1}$.

The evidence for universality summarised here is certainly much weaker than 
for the universality first identified by Feigenbaum\cite{Feig} for 
one-dimensional maps or by Escande and Doveil\cite{Escande} and 
MacKay\cite{MacKay} in 
their renormalisation group analyses of tori, but it is, nevertheless, 
intriguing. Moreover, certain points are seemingly unambiguous. (1) For all 
three Hamiltonians, there is clear evidence for an abrupt change in behaviour 
at or near some critical value ${\epsilon}_{1}$. (2) For 
${\epsilon}>{\epsilon}_{1}$, orbit ensembles evolve towards an escape 
probability 
$P(t)$ which is (a) largely independent of the choice of initial ensemble 
and (b) nearly time-independent, at least for relatively short times. (3) 
The convergence time $T$ depends both on cell size $r$ and 
${\epsilon}-{\epsilon}_{1}$, larger cells and larger 
${\epsilon}-{\epsilon}_{1}$ leading to a more rapid convergence. (4) Because
$T({\epsilon},r)$ increases with decreasing ${\epsilon}-{\epsilon}_{1}$,
pinning down the precise value of ${\epsilon}_{1}$ is quite hard. (5) 
The resulting uncertainties in ${\epsilon}_{1}$ do not impact the apparent
fact that $P$ and $T$ scale in ${\epsilon}-{\epsilon}_{1}$. However, they 
{\it do} impact estimates of the precise values of ${\alpha}$, ${\beta}$, and 
${\delta}$, thus making it difficult to determine whether these exponents are 
the same for all three potentials. What is clear is that, for all three
potentials, the values of the exponents are comparable in magnitude.

\section{
PHASE SPACE FLOW AT LATER TIMES}
\label{sec:level1}
\subsection{Late time evolution of $P(t)$}
\label{sec:level2}
Section II summarised experiments indicating that, on a relatively short time 
scale, the escape probability $P(t)$ evolves towards a near-constant value 
$P_{0}({\epsilon})$. However, there is no reason to expect that $P$ will 
remain approximately constant if the orbit ensembles are evolved for much 
longer times. If, e.g., the initial ensembles contain a few regular orbits 
that cannot escape, these will eventually dominate the orbits that remain in
the system and $P(t)$ must decay towards zero exponentially.
Even if the initial ensembles contain no regular orbits that never escape, one 
might anticipate more complicated behaviour at late times. If, e.g., some 
small subset of the chaotic escape orbits are stuck by cantori near some 
regular island for relatively long times, the escape probability should 
decrease below the initial near-constant $P_{0}$ once the other chaotic orbits
have almost all escaped.

Such a decrease was first noted for orbit ensembles in $H_{2}$ and 
subsequently studied more systematically for ensembles in $H_{3}$ \cite{3}. 
The principal conclusion is that, at least for ${\epsilon}>{\epsilon}_{1}$, 
when orbit ensembles are evolved for somewhat longer times the probability 
$P(t)$ begins to exhibit a slow, monotonic decrease. This is illustrated 
in Fig. 4 which, for one ensemble evolved in $H_{3}$, exhibits $P(t)$ for an
ensemble with ${\sim}{\;}1.1 \times 10^{9}$ orbits carefully chosen from a 
tiny region of
size $r=1 \times 10^{-5}$ to be dominated by ``slow escapers,'' so that 
$P(t)$ can be tracked for a comparatively long interval. 
Visually, $P(t)$ decays too slowly, and has the wrong curvature, to be well 
fit by an exponential. However, 
as is illustrated in Fig. 4b, which plots ${\rm ln}\,P$ as a 
function of ${\rm ln}\,t$, the data for $5<t<180$ or so can be fit quite well 
by a power law, 
$$P(t){\;}{\propto}{\;}t^{-{\mu}}, \eqno(11) $$ 
with ${\mu}=0.39{\pm}0.02.$

This algebraic decay seems very robust, with ${\mu}$ apparently independent 
of the initial cell and, at least within a limited range, the value of 
${\epsilon}$. The best fit value for several different values of ${\epsilon}$ 
(most of which were sampled for shorter times with far fewer orbits) is 
${\mu}=0.35{\pm}0.07$. 

\subsection{Tools of analysis}
\label{sec:level2}

To ascertain why $P(t)$ changes in time and to better understand the 
qualitative character of the flow, orbit ensembles evolved in $H_{3}$ were
also analysed in two other ways.

The first involved computing surfaces of section for an evolving ensemble. 
Each orbit in the ensemble was integrated into the future and, provided that 
it had not yet escaped, its values of $y$ and ${\dot y}$ were recorded at 
successive consequents and sorted to generate sequences of 
surfaces of section exhibiting $(y,{\dot y})$ pairs. 

Such surfaces of section allow one to determine the extent to which the 
ensemble has evolved to cover a large portion of the allowed phase space. 
Moreover, they can facilitate the detection of ``zones of avoidance'' 
associated with regular islands and/or with orbits that have immediately 
escaped, as well as phase space regions with excess concentrations of orbits, 
corresponding presumably to regions from which escape is especially difficult.

The second involved computing short time Lyapunov exponents (cf.
\cite{Grass,Badii,KAA,Voglis}), which probe the average exponential 
instability of 
chaotic orbits over finite time intervals. By analogy with ordinary Lyapunov
exponents, a finite time ${\chi}(t)$ can be defined by the obvious prescription
$${\chi}(t)=\lim_{{\delta}Z(0)\to 0}
{1\over t}\ln {|{\delta}Z(t)|\over |{\delta}Z(0)|} \eqno(12)$$
where 
$|{\delta}Z|^{2}=({\delta}x)^{2}+({\delta}y)^{2}
+({\delta}{\dot x})^{2}+({\delta}{\dot y})^{2}$
denotes the magnitude of the initial phase space perturbation, defined with
respect to the natural Euclidean norm. These exponents were determined 
computationally in the usual way \cite{Bennetin} by introducing a small 
initial perturbation ${\delta}x(0)=
1\times 10^{-12}$ in the $x$-direction and evolving simultaneously 
both the perturbed and unperturbed orbits, periodically renormalising the
amplitude of the perturbation to assure that $|{\delta}Z(t)|$ remains smaller
than $1\times 10^{-8}$. 

At early times, the ${\chi}(t)$ computed in this way will depend strongly on 
the initial perturbation. However, if the orbit segment is chaotic, 
${\chi}(t)$ will quickly become dominated by the component of the perturbation 
in the most unstable direction and become relatively insensitive to the 
initial ${\delta}Z(0)$. Note also that, given ${\chi}(t)$ for two different 
times, $t_{1}$ and $t_{2}$, one can identify the average exponential 
instability for the interval
$t_{1}<t<t_{2}$ as
$$ {\chi}(t_{2}-t_{1})=
{t_{2}{\chi}(t_{2})-t_{1}{\chi}(t_{1})\over t_{2}-t_{1}}.\eqno(13)$$

Short time Lyapunov exponents were used to confirm that, for the values of $h$ 
and ${\epsilon}$ under consideration, most, if not all, of the computed orbits 
in $H_{3}$ are chaotic. Distributions of short time Lyapunov exponents were 
also used to show (1) that the chaotic orbits which have not escaped often 
appear to divide into distinct populations and (2) that there are correlations 
between the magnitude of ${\chi}$ and the time at which the orbit escapes.

\subsection{Surfaces of section}
\label{sec:level2}
Figures 5 a-f exhibit a sequence of sections, generated for one ensemble at 
six different consequents, $t=2$, $6$, $10$, $15$, $20$, and $25$. This 
ensemble was comprised of $160,000$ orbits evolved with ${\epsilon}=1.13$,
a value only marginally above the critical ${\epsilon}_{1}$. The 
cell of initial conditions, with $0<x<0.05$, $0.04<{\dot x}<0.09$, $y=0$, 
$h=1/6$, and hence $0.5681<{\dot y}<0.5780$ is located near the top 
of the energetically accessible portion of the surface of section. The first 
escapes occured at $t=4$, and the largest number of escapes was
at $t=7$. $P(t)$ first settled down towards a smoothly varying form around
$t=10-12$.


Inspection of these, and other intermediate, sections indicates
that the orbits remaining inside the Lyapunov curves tend systematically to 
spread over a relatively large fraction of the energetically allowed phase 
space. Indeed, the elongated striae associated with the specific 
choice of initial conditions, so conspicuous at consequents $t=2$ and $6$, 
have been significantly blurred by $t=10$ and have nearly disappeared by 
$t=25$.

For $t<6-8$, the form of the sections is strongly time-dependent. However, by 
$t=10$ the ensemble has evolved to yield sections characterised by three 
seemingly distinct regions which persist to later times, namely: (1) two 
large holes accompanied by smaller surrounding whorls, (2) several overdense 
regions at positive values of $y$, and (3) a larger region characterised by a 
substantially lower density. As time passes, the occupied regions all decrease
in density, but the overdense regions remain overdense.

The two holes and their surrounding whorls are associated with escapes through 
Lyapunov curves: any orbit with values of $y$ and ${\dot y}$ in these 
regions would already have escaped before intersecting the $x$-axis. For 
larger values of ${\epsilon}$ it is apparent that the visible whorls are part 
of an elaborate set of structures that penetrate throughout large portions of 
the lower density regions, and that this lower density region, which appears 
macroscopically to be populated in a near-uniform fashion, is really laced 
with tiny zones of avoidance.

Assuming that essentially all the orbits in the ensemble are chaotic, the 
overdense regions can be interpreted as reflecting trapping near regular 
islands: Even though these islands may be so small as to be almost 
unobservable, cantori can significantly impact relatively large portions of 
the chaotic phase space\cite{Meiss,LL}. The idea then is that orbits in the 
overdense regions 
are trapped by cantori and can only escape once they have diffused through the 
cantori and can travel unimpeded throughout the remainder of the chaotic sea.

The existence of this three-part structure -- holes, less dense regions, and 
more dense regions -- is independent of the choice of initial 
conditions. Moreover, the general locations of the holes and the higher 
density regions are insensitive to the precise value of ${\epsilon}$. This 
latter fact is manifested in Fig. 6, which shows the analogue of Fig. 5 c, now 
generated for an ensemble with ${\epsilon}=1.06$. This reinforces 
the interpretation that one is seeing the effects of basic phase space 
structures, rather than transient streaming motions reflecting the choice of 
ensemble.
\subsection{Short time Lyapunov exponents}
\label{sec:level2}
Perhaps the most obvious way to search for correlations between the degree of
exponential instability exhibited by a chaotic orbit and the time at which it
escapes is to compute the mean short time Lyapunov exponent, 
${\langle}{\chi}(t_{E}){\rangle}$, for all the orbits in an ensemble that 
escape between successive consequents $t_{E}$ and $t_{E}+1$. The results of one
such computation are presented in Fig. 7, which was generated from an 
ensemble of $1\times 10^{\,6}$ orbits evolved in $H_{3}$ with 
${\epsilon}=1.30$. ${\langle}{\chi}(t_{E}){\rangle}$ clearly exhibits an 
initial relatively rapid decrease for $t_{E}<5-10$, followed by a more 
extended period during which ${\langle}{\chi}(t_{E}){\rangle}$ decreases more 
slowly. (The initial point in Fig. 7 at $t_{E}=5$ {\it is} statistically 
significant.)

As described above, interpreting the computed 
${\langle}{\chi}(t_{E}){\rangle}$ at very early times as an accurate probe of 
the average maximum short time exponent is suspect. Given, however, that the
typical values of ${\chi}$ are greater than or of order unity, the computed 
${\langle}{\chi}(t_{E}){\rangle}$ should be relatively reliable for $t_{E}>5$ 
or so, which means that the initial rapid decrease is most likely real. It is 
not completely clear whether ${\langle}{\chi}(t_{E}){\rangle}$ will continue 
to decrease at late times, or whether it asymptotes towards a nonzero value. 
However, the existence of a continued decrease out to at 
least $t=70$ or so is unquestionably significant statistically.

The observed decrease in ${\langle}{\chi}(t_{E}){\rangle}$ can be easily 
interpreted by assuming that the chaotic phase space inside the 
Lyapunov curves divides into two different regions, namely (1) a region where
orbit segments are less unstable exponentially and from which direct escape to 
infinity is difficult, if not impossible, and (2) a region where orbit segments
are more unstable and from which escape to infinity can proceed on a relatively
short time scale. The idea is that orbits which remain inside the 
Lyapunov curves for a long time will typically spend most of their time in the 
less unstable region before entering the more unstable region and subsequently
escaping to infinity. It follows that, for orbits that only escape at late
times, the short time exponent ${\chi}(t_{E})$, which probes the average
instability for $0<t<t_{E}$, will typically be smaller than for early escapers.


Suppose, oversimplistically, that orbit segments in the less and more unstable
regions can be characterised respectively by unique short time exponents
${\chi}_{L}$ and ${\chi}_{H}$, and that any orbit that enters the high
${\chi}$ region will escape after exactly a time $t_{e}$. It then follows 
from eq. (13) 
that the total ${\chi}(t_{E})$ for an orbit escaping at time $t_{E}$ satisfies
$${\chi}(t_{E})={1\over t_{E}}{\Bigl[}t_{e}\,{\chi}_{H} 
+ (t_{E}-t_{e})\,{\chi}_{L}{\Bigr]}. \eqno(14)$$
Consistent with Fig.~7, this implies that ${\chi}(t_{E})$ decreases 
monotonically, but eventually asymptotes towards a nonzero ${\chi}_{L}$. For 
$t_{E}>15$ or so the computed ${\langle}{\chi}(t_{E}){\rangle}$ exhibited 
in Fig. 7 is in fact well fit by eq. (14) with ${\chi}_{L}=0.92$ and 
$t_{e}({\chi}_{H}-{\chi}_{L})=7.80$. 


To confirm that the observed decrease in ${\langle}{\chi}(t_{E}){\rangle}$ 
reflects transitions between relatively distinct orbit 
populations, it is also instructive to determine how short time Lyapunov 
exponents computed for the same set of orbits change as a function of time. 
One way to do this is to consider all the orbits in an initial ensemble that
escape at a given time $T_{E}$ and, given expressions for ${\chi}(t)$ for
times $t<T_{E}$, computed using eq.~(12), extract short time exponents for
different intervals $t_{1}<t<t_{2}$.

Figure 8, generated from the same ensemble as Fig.~7, focuses on all
$832$ orbits that escaped at $t=50$, computing distributions of short time 
exponents for the intervals $15<t<20$ (solid curve) and $45<t<50$ (dashed 
curve). 
Both distributions are
bimodal, seemingly comprised of two different populations with peaks at or 
near the same values of ${\chi}$, but it is clear that the relative
height of the two peaks changes significantly in time. For the earlier 
interval, the lower ${\chi}$ population dominates whereas for the later 
interval the low and high ${\chi}$ populations are of more nearly equal
importance. For $10<t<15$ the mean ${\langle}{\chi}{\rangle}=0.70$; for 
$45<t<50$ the mean ${\langle}{\chi}{\rangle}=1.11$. This is consistent with 
the interpretation that most of the orbits that escaped at $t=50$ were members 
of a low ${\chi}$ population at early times but shifted to the higher ${\chi}$ 
population shortly before escaping from the system.

\section{PHYSICAL INTERPRETATION}
\label{sec:level1}
\subsection{General considerations}
\label{sec:level2}
This section suggests a simple model for the behaviour observed in Sections II
and III which is based on the assumption that, over finite intervals, orbits
divide at least approximately into three distinct classes, namely (1) regular 
orbits, (2) sticky, or temporarily confined, 
orbits, and (3) unconfined chaotic orbits, even though the distinction between
confined and unconfined chaos disappears entirely in the $t\to\infty $ limit.

Conventional wisdom would suggest that the presence of stable periodic orbits, 
which one expects for generic Hamiltonians, implies that there must exist a
finite measure of regular orbits, even at very high energies $h$ and/or large 
values of ${\epsilon}$. However, the existence of such regular regions, 
separated from the remaining chaotic orbits by invariant {\it KAM} tori, 
suggests in turn that the surrounding chaotic sea should contain cantori which,
albeit not impenetrable barriers, can trap chaotic orbits near the regular 
islands for relatively long intervals of time. Even if there is no absolute 
distinction between different types of orbits in the chaotic sea, there may 
exist short time {\it de facto} distinctions which can have significant 
implications for the Hamiltonian flow. (Strictly speaking, in general there 
will also exist a finite measure of chaotic orbits {\it inside} the 
{\it KAM} tori. However, these can never breach the invariant tori and, as
such, will never be able to escape to enter the surrounding stochastic sea.
For this reason, they may be lumped together with the regular orbits in the
following discussion.)

If a localised ensemble of phase space points, each corresponding to a chaotic
orbit with energy $h$, is evolved in a time-independent potential which, as
for the Hamiltonians (1) - (3) for ${\epsilon}<{\epsilon}_{0}$, has a compact
constant energy hypersurface, one anticipates a coarse-grained evolution 
towards an invariant distribution corresponding to a uniform population of the
accessible phase space\cite{LL}. Numerical experiments suggest (cf. 
\cite{KPRE}) that, if the chaotic phase space is not significantly impacted by 
cantori, so that a single chaotic orbit can easily access the entire region 
without having to diffuse through any barriers (cf. \cite{MMP1,MMP2}), 
this approach will proceed exponentially in time on a time scale of order the
natural crossing time, $t_{cr}$. If, however, cantori play an important role 
in partitioning the chaotic phase space regions over relatively short time 
scales, one can instead observe a more complex, seemingly two stage process 
\cite{MMP1,MMP2,MABK}. Sets of orbits in different nearly disjoint regions 
will 
rapidly approach near-invariant distributions, corresponding to near-uniform 
populations of the separate regions; but only later, on a significantly longer 
time scale, will orbits from different regions ``mix'' to yield an approach 
towards a true invariant distribution.

Suppose now that ${\epsilon}>{\epsilon}_{0}$, so that orbits are no longer 
bound energetically. It then seems reasonable to interpret the observed 
behaviour of orbits in terms of two distinct sorts of ``escape,'' 
namely (1) unconfined chaotic orbits which pass through Lyapunov curves to 
escape to infinity and (2) confined chaotic orbits, originally stuck near the 
regular regions, which pass through one or more cantori to become unconfined, 
after which they too can escape to infinity. To the extent that the escape 
channels -- both the simple gaps breached by Lyapunov curves and the 
more complex cantor sets of holes in cantori -- are small, or that one is 
considering phase space regions relatively far from the escape channels, it 
should be reasonable to visualise what is happening in terms of ensembles that 
have evolved towards a near-invariant distribution. 

Suppose, in particular, that one selects an initial ensemble where most of the 
orbits are unconfined chaotic, but that there are also a significant number of 
confined chaotic orbits and, perhaps, a few regular orbits. It is then easy to 
explain the qualitative evolution.

On a relatively short time scale, the unconfined chaotic orbits should evolve 
towards a near-uniform population of the phase space regions far from the 
Lyapunov curves and outside any cantori which significantly impede phase space 
transport. (The fact that, in the late time sections of Fig. 5, different 
parts of the lower density region seem to have the same relative density at 
different times supports this idea.) However, once this near-invariant 
distribution has been achieved, escapes to infinity should proceed ``at 
random'' at a near-constant rate, so that the unconfined chaotic orbits will 
be characterised by a constant escape probability. To the extent that the 
total orbit population is dominated by the initially unconfined orbits, and 
that appreciable numbers of confined orbits have not yet become unconfined, 
the total escape probability $P$ should be approximately constant.

Eventually, however, most of the unconfined chaotic orbits will have escaped,
so that, assuming that unconfined orbits cross the Lyapunov curves 
more quickly than confined orbits diffuse through cantori, the total escape 
probability is impacted significantly, and ultimately dominated, by the 
remaining orbits. If all these orbits were regular and unable to escape, one 
would expect $P$ to decay exponentially in time (cf. \cite{1}). Given, however,
that most of the remaining orbits are chaotic orbits originally trapped by 
cantori, as seems true for ${\epsilon}>{\epsilon}_{1}$, one expects a slower 
decay in $P$ reflecting transitions from confined to unconfined chaos: The 
idea here is that confined orbits will only diffuse through cantori to become 
unconfined very slowly, but that, once unconfined, they will quickly escape to 
infinity. The observed $P{\;}{\propto}{\;}t^{-\mu}$ is thus driven by the 
diffusion of orbits through cantori rather than their escape through 
Lyapunov curves. That diffusion through cantori need not characterised by a 
constant escape probability is in fact well known (cf. \cite{Meiss}).

The same qualitative picture should remain valid even if the initial 
orbit ensemble is carefully selected to contain no chaotic orbits 
trapped near the regular regions. Even if most of the original unconfined 
orbits quickly escape to infinity through one of the Lyapunov curves, a 
small fraction of those orbits could become trapped by cantori. However, 
once trapped most of these orbits will only leak out at significantly later 
times, when the population of unconfined orbits inside the Lyapunov curves 
has become significantly reduced.

\subsection{A simple model}
\label{sec:level2}
Consider an ensemble of initial conditions of energy $h$, located inside the 
Lyapunov curves, which may be divided into three different classes --  
unconfined chaotic, temporarily confined chaotic, and regular -- characterised 
by numbers $N_{u}(0){\;}{\gg}{\;}{\hat N}_{c}(0){\;}{\gg}{\;}N_{r}(0)$. Now 
implement a probabilistic description, treating the regular orbits as a 
separate population that can never escape, but allowing for three sorts 
of transitions, namely escapes of unconfined orbits through Lyapunov curves, 
trapping of unconfined orbits by cantori, and leakage of confined orbits to 
become unconfined orbits inside the Lyapunov curves. 

Calculating the total escape rate $R$, the continuum limit of the discrete
probability $P$ computed in the numerical experiments summarised above, 
is straightforward if one makes two basic assumptions, each  implicit in 
the preceding and seemingly consistent with the numerical experiments. (1) The 
rate at which unconfined orbits escape to infinity assumes a constant value 
${\lambda}$, independent of time. (2) ${\lambda}$ is much larger than the 
rates at which unconfined orbits become temporarily confined and temporarily 
confined orbits become unconfined. 

These assumptions imply that, at early times, the total escape rate is 
dominated by the escape of initially unconfined orbits, and that
the details of any early trapping of unconfined orbits are irrelevant. This
means that, when computing the total escape rate, the confined population may 
be approximated by an expression of the form
$$N_{c}(t)=N_{c}(0)f(t), \eqno(15)$$
where $N_{c}(0)$ allows for a possible early time trapping of some small
fraction of the originally unconfined orbits and 
$f(t)$ is a monotonically decreasing function of time, assumed to 
satisfy an initial condition $f(0)=1$. The rate $R_{cu}$ at which temporarily 
confined orbits become unconfined is thus
$$R_{cu}(t)=-{1\over f}{df\over dt}  \eqno(16). $$
Similarly, $N_{u}(t)$ must satisfy a simple rate equation
$$\hskip -.4in {dN_{u}(t)\over dt}={dN_{u}(t)\over dt}{\Biggl|}_{out}+
{dN_{u}(t)\over dt}{\Biggl|}_{in}$$
$$=-{\lambda}N_{u}(t)-N_{c}(0){df(t)\over dt}. \eqno(17) $$

This latter equation is easily solved to yield 
$$N_{u}(t)=N_{u}(0)e^{-{\lambda}t}-N_{c}(0)\,\int_{0}^{t}\,
d{\tau}e^{{\lambda}({\tau}-t)}\;{df({\tau})\over d{\tau}} \;.\eqno(18) $$
However, this implies that the total escape rate
$$R(t){\;}{\equiv}{\;}{-1\over (N_{r}+N_{c}+N_{u})}{\;}{d\over dt}
(N_{r}+N_{c}+N_{u}) $$
$$={{\lambda}{\Bigl[}1-{\nu}_{c}\,\int_{0}^{t}\,d{\tau}
e^{{\lambda}{\tau}}df({\tau})/d{\tau}{\Bigr]}\over
{\Bigl[}1-{\nu}_{c} \,\int_{0}^{t}\,d{\tau}
e^{{\lambda}{\tau}}df({\tau})/d{\tau}+{\nu}_{c}f(t)e^{{\lambda}t}+{\nu}_{r}
e^{{\lambda}t}{\Bigr]}}\;, \eqno(19)$$
\vskip .1in
\par\noindent
where
${\nu}_{c}=N_{c}(0)/N_{u}(0)$ and ${\nu}_{r}=N_{r}/N_{u}(0)$ reflect
original relative abundances. 

That the escape rate through the Lyapunov curves in much larger than the 
rate at which confined orbits become unconfined means that 
${\lambda}{\;}{\gg}{\;}|(1/f)(df/dt)|.$
This implies, however, that the integrals in the preceding equation can be
evaluated perturbatively, expanding $df({\tau})/d{\tau}$ about its value at 
time ${\tau}=t$. Recognising that the second term in the denominator is 
small compared with the third, one thus concludes that 
$$R(t){\;}{\approx}{\;}{{\lambda}-{\nu}_{c}e^{{\lambda}t}df(t)/dt\over
1+{\nu}_{c}e^{{\lambda}t}f(t)+{\nu}_{r}e^{{\lambda}t}}\; . \eqno(20)$$

Granted that ${\nu}_{r}{\;}{\ll}{\;}{\nu}_{c}{\;}{\ll}{\;}1$, $R(t)$ typically
admits three different, asymptotic regimes, namely
\par\noindent
(1) an early time regime where $R{\;}{\approx}{\;}{\lambda}$;
\par\noindent
(2) an intermediate regime where $R \to
-(1/f)df/dt$; and
\par\noindent
(3) a late time 
regime where $R \to -({\nu}_{c}/{\nu}_{r})df/dt$.
\par\noindent
At early times, the total escape rate is fixed by the rate at which initially 
unconfined orbits cross the Lyapunov curves. Later, once most of these 
original unconfined orbits have escaped, $R(t)$ is set by the rate at 
which confined chaotic orbits become unconfined. Finally, once most of the 
initially confined orbits have escaped, there is a more rapid decrease in $R$ 
towards zero, reflecting the fact that the now dominant regular population can 
never escape.

It is natural to interpret the probability $P{\;}{\propto}{\;}t^{-{\mu}}$ 
described in Section III as reflecting the intermediate regime. If ${\mu}$ 
were equal to unity, one would then infer a population 
$$N_{c}(t){\;}{\propto}{\;}t^{-q}, \eqno(21)$$
for some constant $q>0$, i.e., an eventual power law decay in the number of
confined orbits. Given, however, that the best fit ${\mu}<1$, one infers 
instead a population
$$N_{c}(t){\;}{\propto}{\;}\exp (-qt^{1-\mu}), \eqno(22)$$
which decays faster than an a power law but slower than the exponential
decrease associated with a constant escape rate. 

That $P(t)$ is not constant is hardly surprising. Indeed, diffusion through
cantori, interpreted as orbits wending their way through a self-similar
collection of turnstiles,\cite{MMP1,MMP2} would suggest that $P(t)$ decay 
in time (cf. 
\cite{Chirikov,Karney,Hanson}). Rather, what is interesting is that 
${\mu}{\;}{\ne}{\;}1$.
In the context of chaotic scattering, one seems to see a sharp distinction 
between hyperbolic scattering (cf. \cite{Ott2,Smil}), where the number of 
incident
particles remaining within the scattering region at time $t$ decays
exponentially, i.e., $N(t){\;}{\propto}{\;}\exp (-{\lambda}t)$, and 
nonhyperbolic scattering (cf. \cite{Yau}), where the number remaining decays 
as a power law, i.e., $N(t){\;}{\propto}{\;}t^{-q}$. The origin of the 
intermediate behaviour observed here is, at the present, unclear.


This picture presupposes that 
${\nu}_{r}{\;}{\ll}{\;}{\nu}_{c}{\;}{\ll}{\;}1$. Indeed, if these 
inequalities fail, the qualitative evolution can change significantly. 
Suppose, for instance, that temporarily confined chaotic orbits are 
unimportant compared with regular 
orbits, so that one need consider only two orbit classes -- regular and 
unconfined --, and that, even early on, there are many more regular than
unconfined orbits. It then follows immediately that, already at early times,
$R(t)$ should decay towards zero exponentially:
$$R(t){\;}{\approx}{\;}{{\lambda}\over 1 + (N_{r}/N_{c}){\exp}(-{\lambda}t)}
{\;}\to{\;}{N_{c}\over N_{r}}{\,}{\exp}(-{\lambda}t) .\eqno(23)$$
This is consistent with the observed behaviour for 
${\epsilon}_{0}<{\epsilon}<{\epsilon}_{1}$ where, as noted already, $P(t)$ 
decays towards zero without first asymptoting
towards a near constant nonzero value. 
\subsection{Discussion}
\label{sec:level2}
To place the preceding in an appropriate context, three important
points should be noted.
\par\noindent
1. Extracting a near-constant $P_{0}({\epsilon})$ is 
necessarily a somewhat imprecise operation. The algorithm described in Paper 1,
or any obvious alternative, depends crucially on the idea that $P(t)$ 
will evolve towards a form that is largely independent of the initial 
conditions on a time scale sufficiently short that appreciable numbers of 
sticky chaotic orbits do not escape and the total escape probability is 
dominated by the near-constant rate at which unconfined orbits escape. If the 
ensemble ``forgets'' its initial conditions sufficiently quickly, it is 
possible operationally to identify a reasonable estimate of  
$P_{0}({\epsilon})$. Strictly speaking, however, 
the ensemble is really evolving towards a characteristic $P({\epsilon},t)$
which manifests a slow, but nontrivial, time-dependence. This leads to an 
inherent inaccuracy in the determination of the values of the critical
${\epsilon}_{1}$ and, especially, the exponents ${\alpha}$, ${\beta}$, and
${\delta}$.
\par\noindent
2. The calculations described in Section III were restricted to relatively 
short times, $t<200$ or so; and, for this reason, one cannot preclude the 
possibility that, on a significantly longer time scale, $P(t)$ could assume
a form very different from what was suggested in Section IV. For example, 
one cannot preclude the possibility that, for much later times, 
$P(t){\;}{\propto}{\;}t^{-1}$, in agreement with Karney's \cite{Karney} 
experiments. Ideally one might like to integrate for much longer times, but 
this quickly becomes very expensive computationally: because the overwhelming 
majority of the orbits in an initial ensemble escape relatively early on, one 
would need to start with an absolutely enormous collection of initial 
conditions in order to derive statistically significant conclusions about 
behaviour at much later times. Nevertheless, even though one cannot exclude 
the possibility of different later time behaviour, it is significant that the 
observed scaling  $P(t){\;}{\propto}{\;}t^{-\mu}$ for $t<180$ or so appears to 
be robust
\par\noindent
It should also be noted that one cannot completely exclude the possibility that
the long time integrations described in Section III are contaminated by 
accumulating errors which, e.g., make the system slightly dissipative. 
Numerical tests described in Paper I allow one to be confident that the early 
time computations ($t<20$ or so) are reliable, but there is less hard evidence
to justify accepting long time integrations at face value. All that can be said
definitively is that, even for the longest time integrations, the relative
energy error for an orbit was never larger than $2\times 10^{-8}$ and usually
orders of magnitude smaller.
\par\noindent
3. This simple three-component model, based on two, and only two, nearly 
distinct classes of chaotic orbits, may well be oversimplistic. 
Indeed, lumping together every chaotic orbit that is not unconfined into a
single population assumed to have reached a statistical near-equilibrium seems 
less justified than assuming that, at least away from the Lyapunov curves, the 
unconfined phase space is characterised by a near-invariant distribution.
Furthermore, the 
entire analysis assumes an abrupt transition occuring at or near some 
critical ${\epsilon}_{1}$. However, one can argue that these limitations are
not completely unreasonable. Detailed examinations of orbit ensembles 
on a compact phase space hypersurface indicate that, oftentimes, many of the 
qualitative features of a flow can be interpreted by allowing only for two
classes of chaotic orbits, named sticky and unconfined (cf. 
\cite{MABK,MMP1,MMP2}). Moreover, 
investigations of the effects of increasing deviations from integrability 
suggest that, as the control parameter ${\epsilon}$ becomes larger, holes in
cantori can abruptly increase in size, so that what was initially a barrier
that could only be penetrated on a very long time scale ceases to play a
significant role in impeding phase space transport \cite{CVB}.

\acknowledgments
H. E. K. acknowledges useful discussions with Elaine Mahon and Ed Ott
regarding short time Lyapunov exponents and the problem of convergence 
towards invariant and near-invariant distributions. 
H. E. K. was supported in part by the NSF grant PHY92-03333. 
The remaining authors were supported in part by the European Community Human 
Capital and Mobility Program ERB4050 PL930312.
Some of the computations in this paper were effected using computer time 
made available through the Research Computing Initiative at the Northeast 
Regional Data Center (Florida) by arrangement with IBM.

\vskip 1in
 \begin{table}
\caption{Critical escape parameters
\label{table1}}
\vskip .1in
\begin{tabular}{llll}
&&& \\
& $H_{1}$ & $H_{2}$ & $H_{3}$ \\
&&& \\
\tableline
&&& \\
$h$ & $0.12$ & $0.125$ & $1/6{\;}{\approx}{\;}0.167$ \\
${\epsilon}_{0}$ & $1/(4h){\;}{\approx}{\;}2.08$ & $1/\sqrt{8h} =1.00$ & $1.00$
\\
${\epsilon}_{1}$ & $4.90\pm0.01$ & $1.15^{+0.02}_{-0.05}$ & $1.10{\pm}0.05$ \\
${\alpha}$ & $0.46{\pm}0.05$ & $0.46{\pm}0.05$ & $0.45{\pm}0.05$ \\
${\beta}$ & $0.39^{+0.14}_{-0.06}$ & $0.50^{+0.15}_{-0.10}$ & 
$0.37^{+0.10}_{-0.07}$ \\
${\delta}$ & $0.08{\pm}0.03$ & $0.11{\pm}0.03$ & $0.12{\pm}0.03$ \\
&&& \\
\end{tabular}
\end{table}

\eject

\pagestyle{empty}
\begin{figure}[t]
\centering
\centerline{
        \epsfxsize=5 cm
        \epsffile{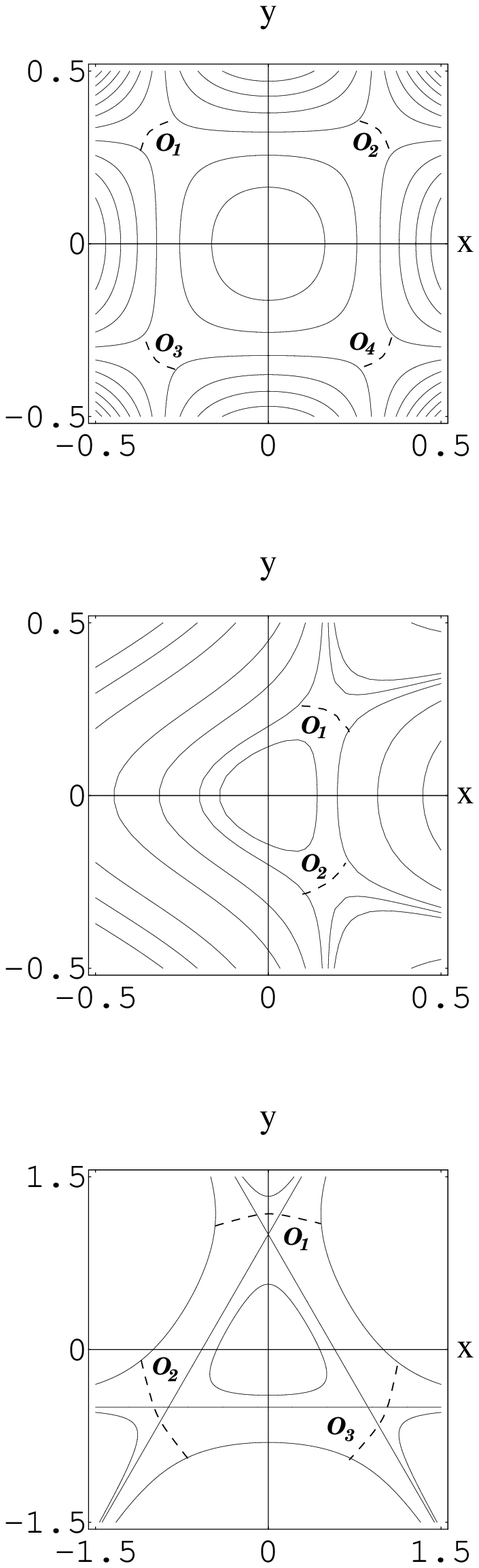}
           }
        \begin{minipage}{10cm}
        \end{minipage}
        \vskip -0.0in\hskip -0.0in

\caption{Equipotential surfaces of the potential for (a) $H_{1}$ with
${\epsilon}=5.26$, (b) $H_{2}$ with ${\epsilon}=3.0$, and (c) $H_{3}$ with
${\epsilon}=1.0$. $O_{1}$, $O_{2}$, $O_{3}$ and $O_{4}$ represent Lyapunov
unstable periodic orbits.}

\vspace{-5.0cm}
\end{figure}
\vfill\eject

\pagestyle{empty}
\begin{figure}[t]
\centering
\centerline{
        \epsfxsize=8 cm
        \epsffile{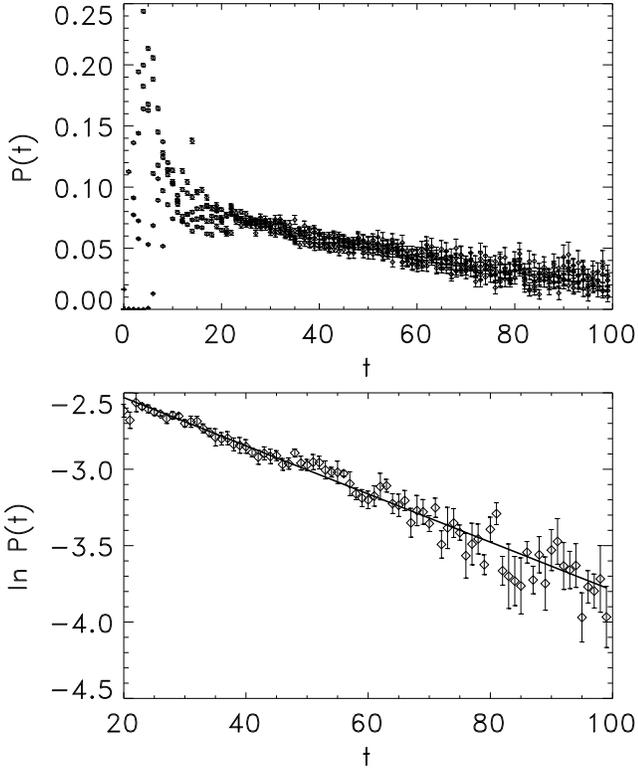}
           }
        \begin{minipage}{10cm}
        \end{minipage}
        \vskip -0.0in\hskip -0.0in

\caption{ (a) The escape probability, $P(t)$, computed for five different 
cells of initial conditions evolved in $H_{2}$ with 
${\epsilon}=1.04<{\epsilon}_{1}$. The solid curve is an exponential fitted to 
the interval $20<t<100$. The error bars reflect uncertainties computed as in
eq. (5). (b) The same data on a log-log plot. Here the orbits from the 
different cells have been combined to yield a single $P$ and ${\Delta}P$,
again computed as in eq. (5).}
\vspace{-5.0cm}
\end{figure}
\vfill\eject

\pagestyle{empty}
\begin{figure}[t]
\centering
\centerline{
        \epsfxsize=8 cm
        \epsffile{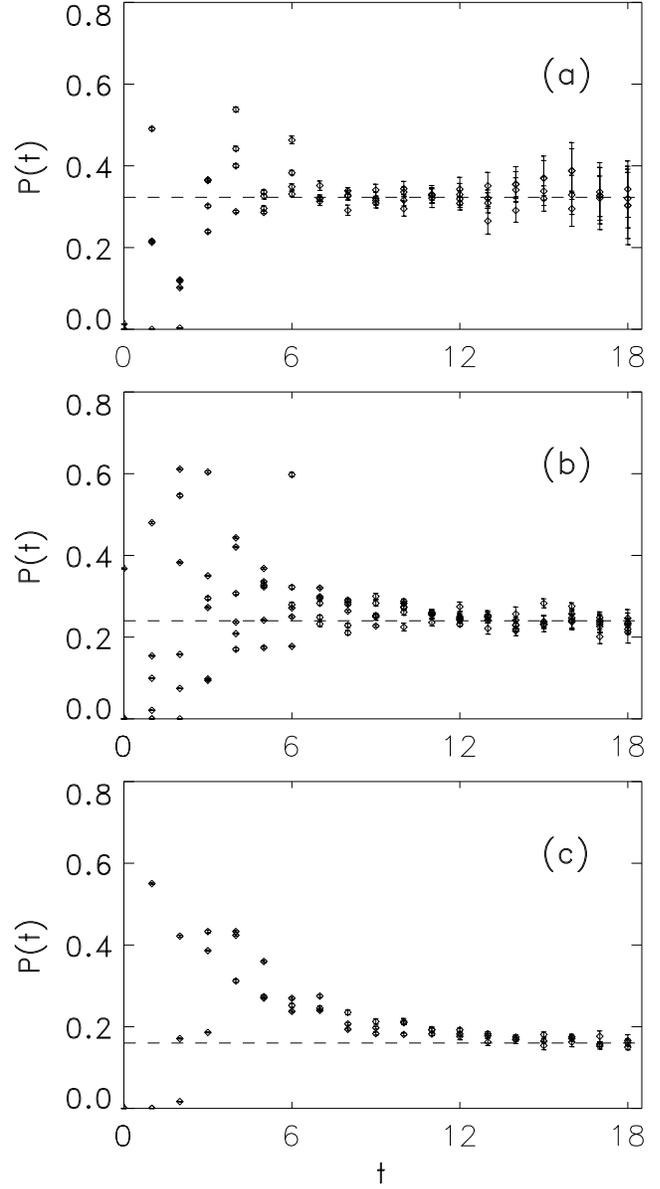}
           }
        \begin{minipage}{10cm}
        \end{minipage}
        \vskip -0.0in\hskip -0.0in

\caption{ (a) The escape probability, $P(t)$, computed for several different 
cells of initial conditions in $H_{1}$ with ${\epsilon}=5.05>{\epsilon}_{1}$. 
The dashed line indicates the best fit $P_{0}$.
(b) The same for $H_{2}$ with ${\epsilon}=1.20$.
(c) The same for $H_{3}$ with ${\epsilon}=1.30$}
\vspace{-5.0cm}
\end{figure}
\vfill\eject

\pagestyle{empty}
\begin{figure}[t]
\centering
\centerline{
        \epsfxsize=8 cm
        \epsffile{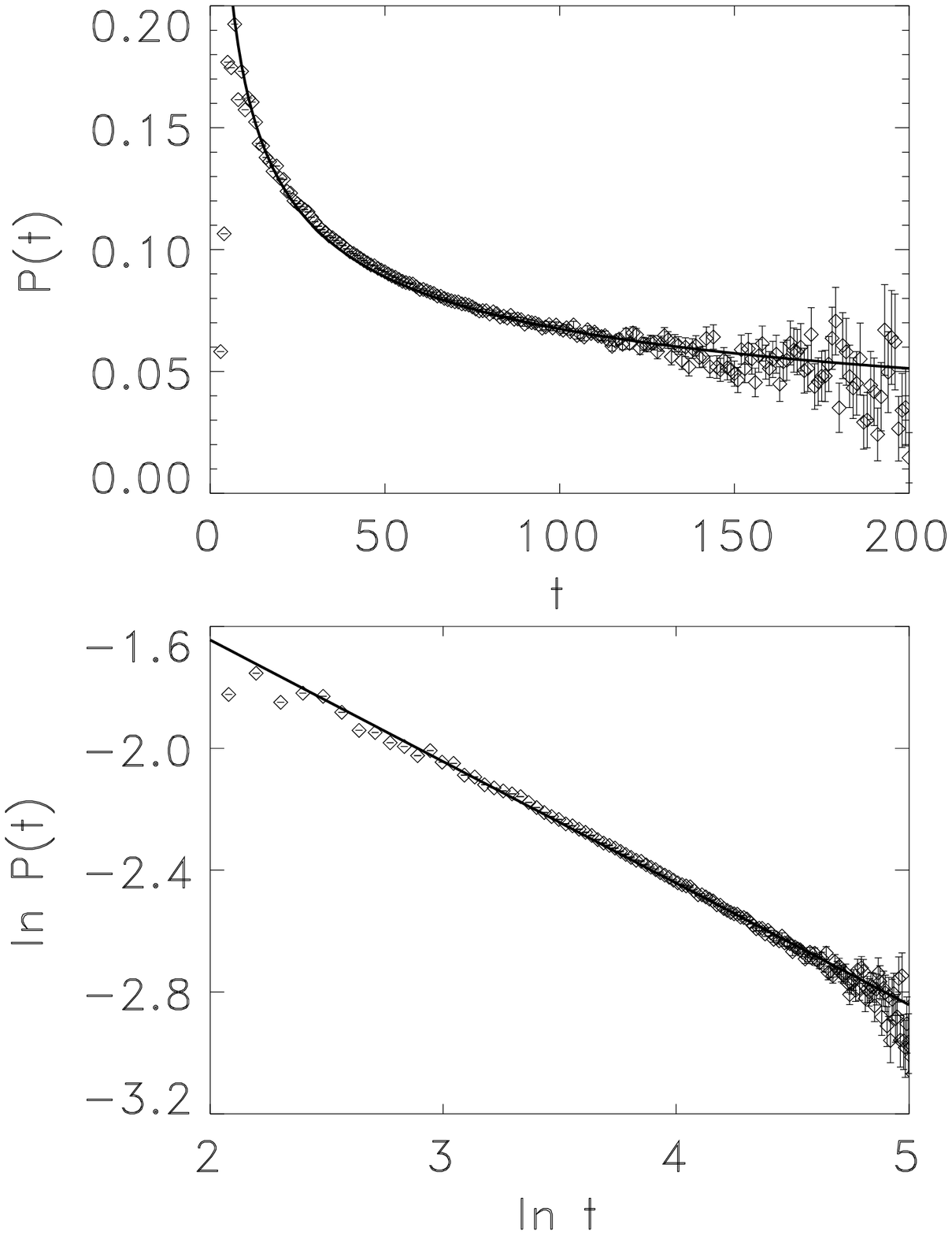}
           }
        \begin{minipage}{10cm}
        \end{minipage}
        \vskip -0.0in\hskip -0.0in

\caption{ (a) The escape probability, $P(t)$, computed for the same initial 
conditions as in Fig. 3 (c), now allowing for much longer times. The solid 
curve exhibits a power law fit, $P{\;}{\propto}{\;}t^{-\mu}$, with 
${\mu}=0.39.$ (b) The same data on a semilog plot, analysed as for Fig. 2b.}
\vspace{-0.0cm}
\end{figure}
\vfill\eject
.\vfill\eject
\pagestyle{empty}
\begin{figure}[t]
\centering
\centerline{
        \epsfxsize=6 cm
        \epsffile{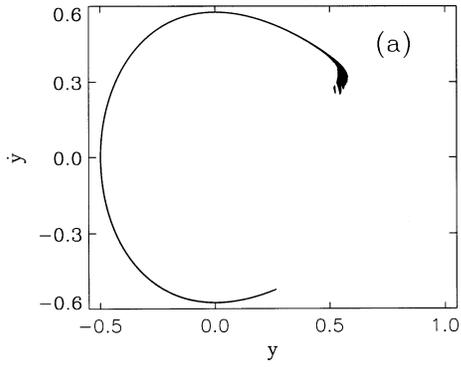}
           }
        \begin{minipage}{10cm}
        \end{minipage}
        \vskip -0.0in\hskip -0.0in

\caption{ (a) Surface of section at $t=2$ for an ensemble of initial 
conditions evolved in $H_{3}$ with ${\epsilon}=1.13$, exhibiting $y$ and 
$\dot y$ for every orbit still inside the Lyapunov curves that crosses the 
$x=0$ axis with ${\dot x}<0$ between consequents $t$ and $t+1$.  }
\vspace{-0.0cm}
\end{figure}

\pagestyle{empty}
\begin{figure}[t]
\centering
\centerline{
        \epsfxsize=6 cm
        \epsffile{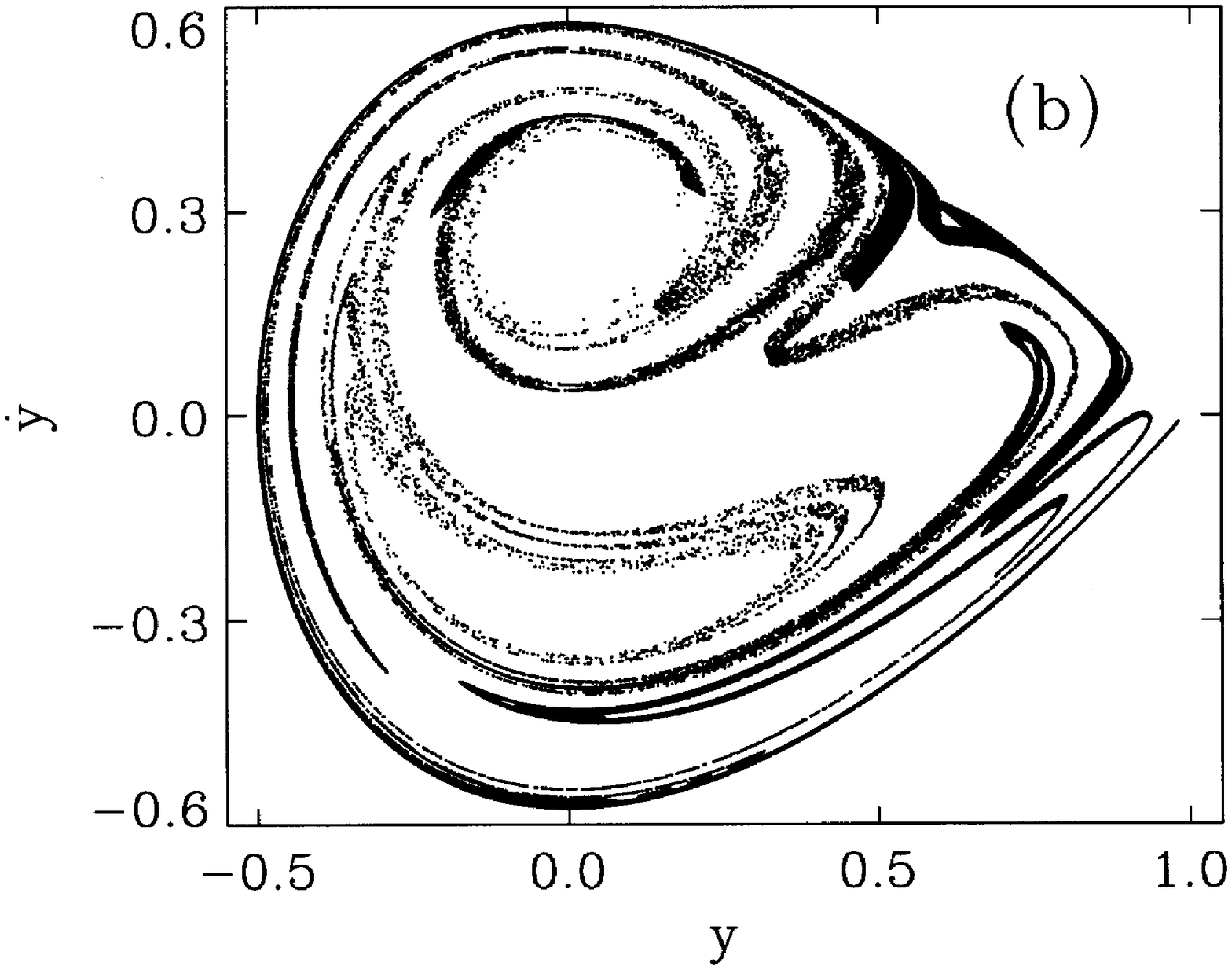}
           }
        \begin{minipage}{10cm}
        \end{minipage}
        \vskip -0.0in\hskip -0.0in

(b) The same for $t=6$.
\vspace{-0.0cm}
\end{figure}

\pagestyle{empty}
\begin{figure}[t]
\centering
\centerline{
        \epsfxsize=6 cm
        \epsffile{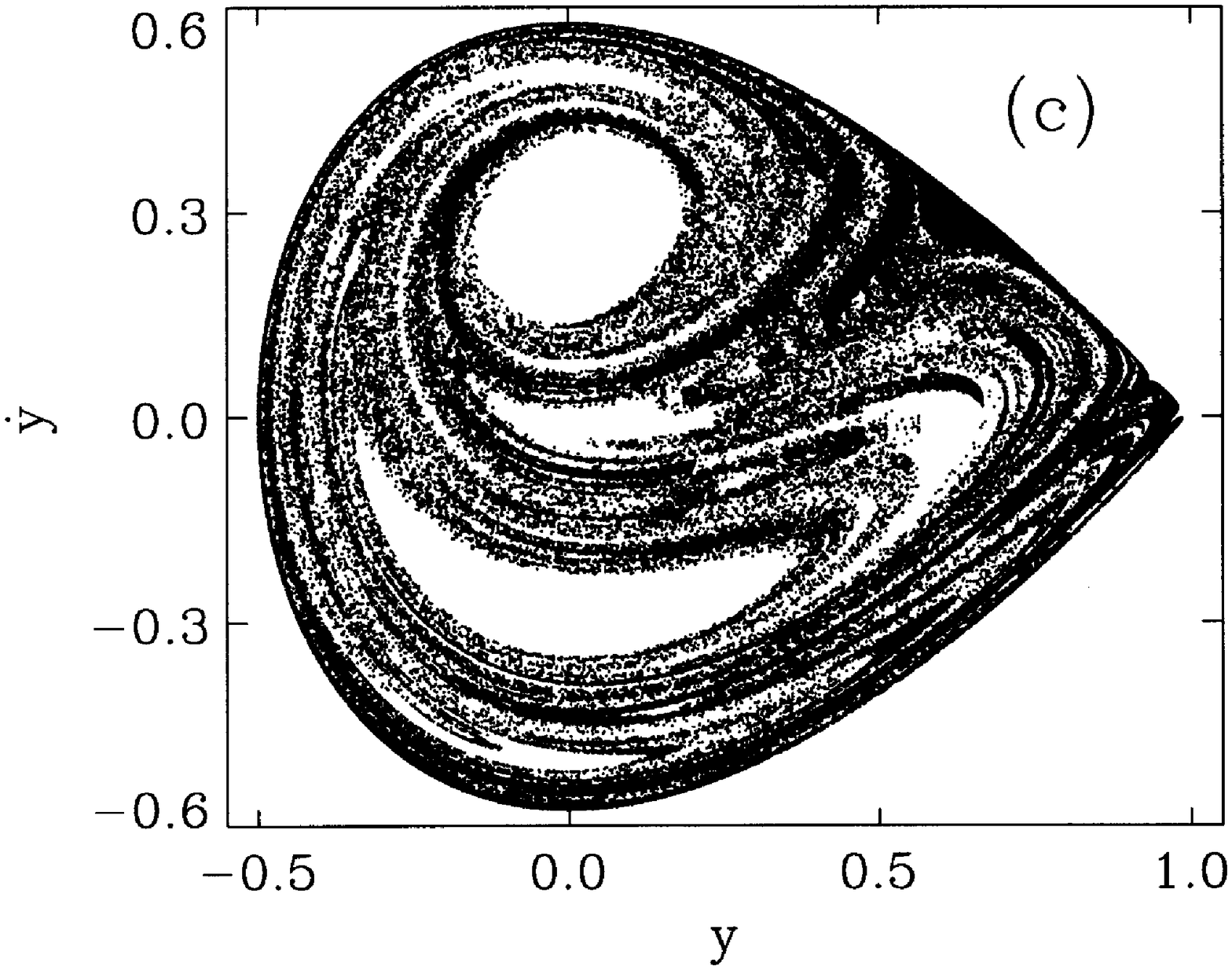}
           }
        \begin{minipage}{10cm}
        \end{minipage}
        \vskip -0.0in\hskip -0.0in

(c) The same for $t=10$.
\vspace{-0.0cm}
\end{figure}
\vfill\eject

\pagestyle{empty}
\begin{figure}[t]
\centering
\centerline{
        \epsfxsize=6 cm
        \epsffile{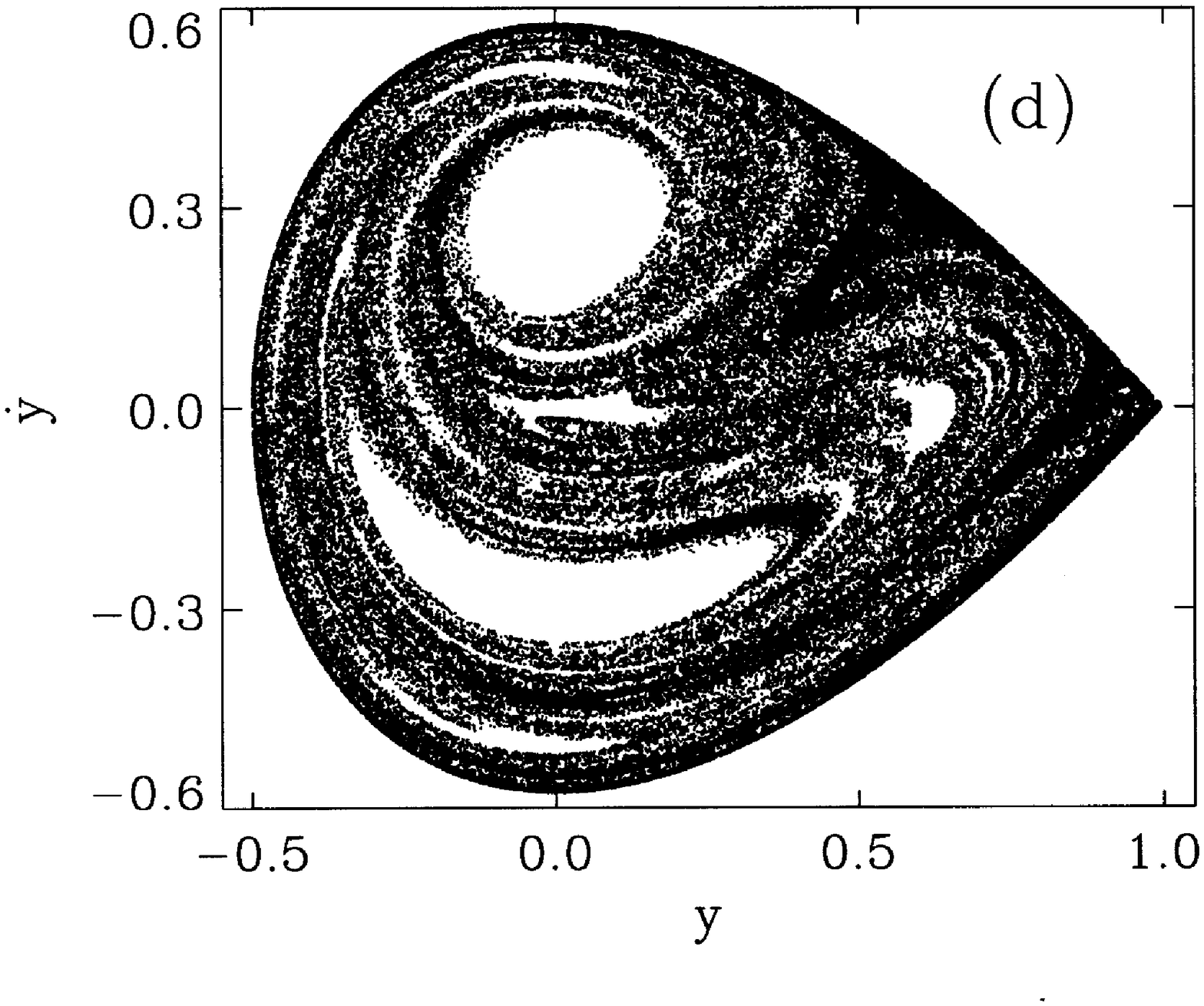}
           }
        \begin{minipage}{10cm}
        \end{minipage}
        \vskip -0.0in\hskip -0.0in

(d) The same for $t=15$.
\vspace{-0.0cm}
\end{figure}

\pagestyle{empty}
\begin{figure}[t]
\centering
\centerline{
        \epsfxsize=6 cm
        \epsffile{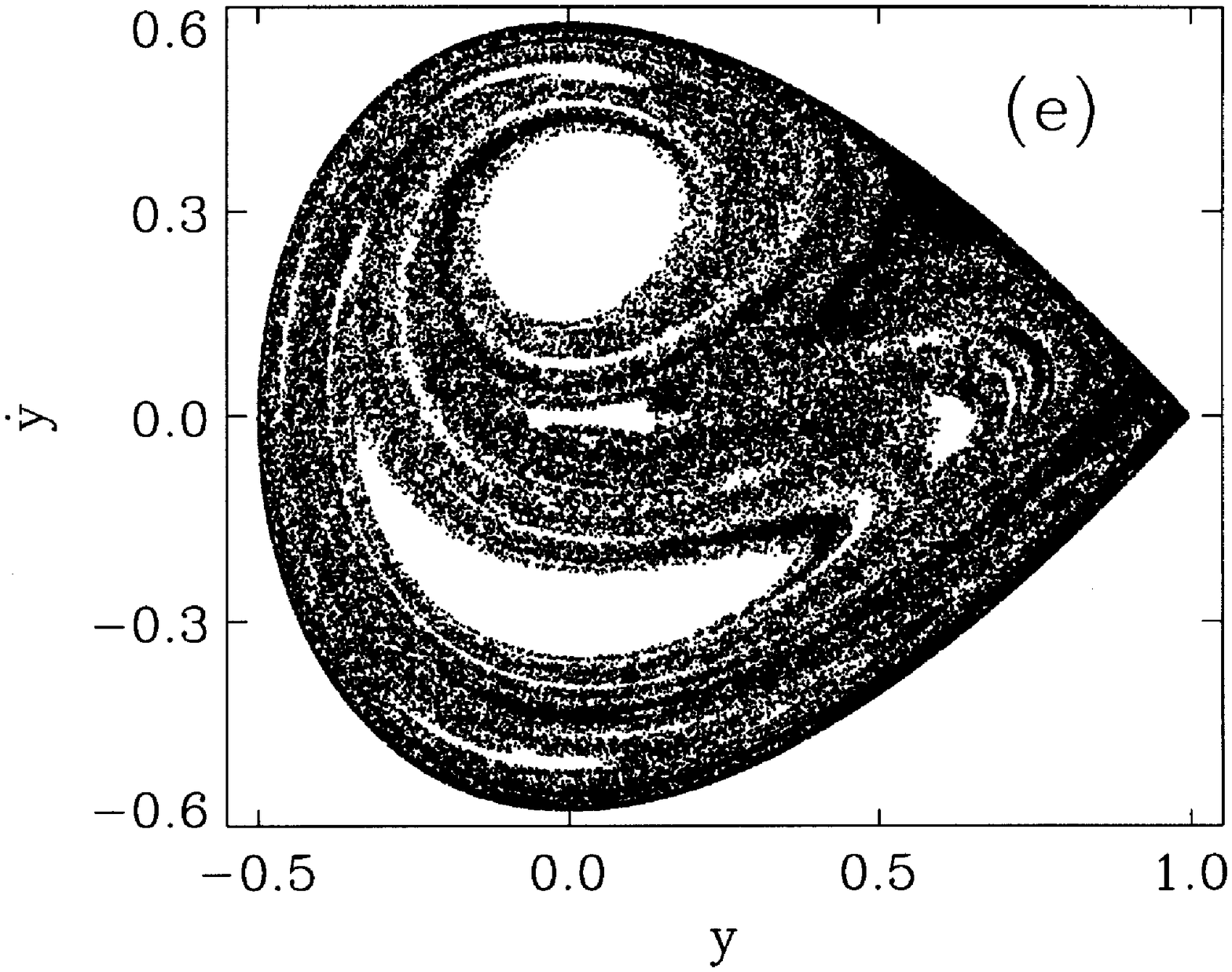}
           }
        \begin{minipage}{10cm}
        \end{minipage}
        \vskip -0.0in\hskip -0.0in

(e) The same for $t=20$.
\vspace{-0.0cm}
\end{figure}

\pagestyle{empty}
\begin{figure}[t]
\centering
\centerline{
        \epsfxsize=6 cm
        \epsffile{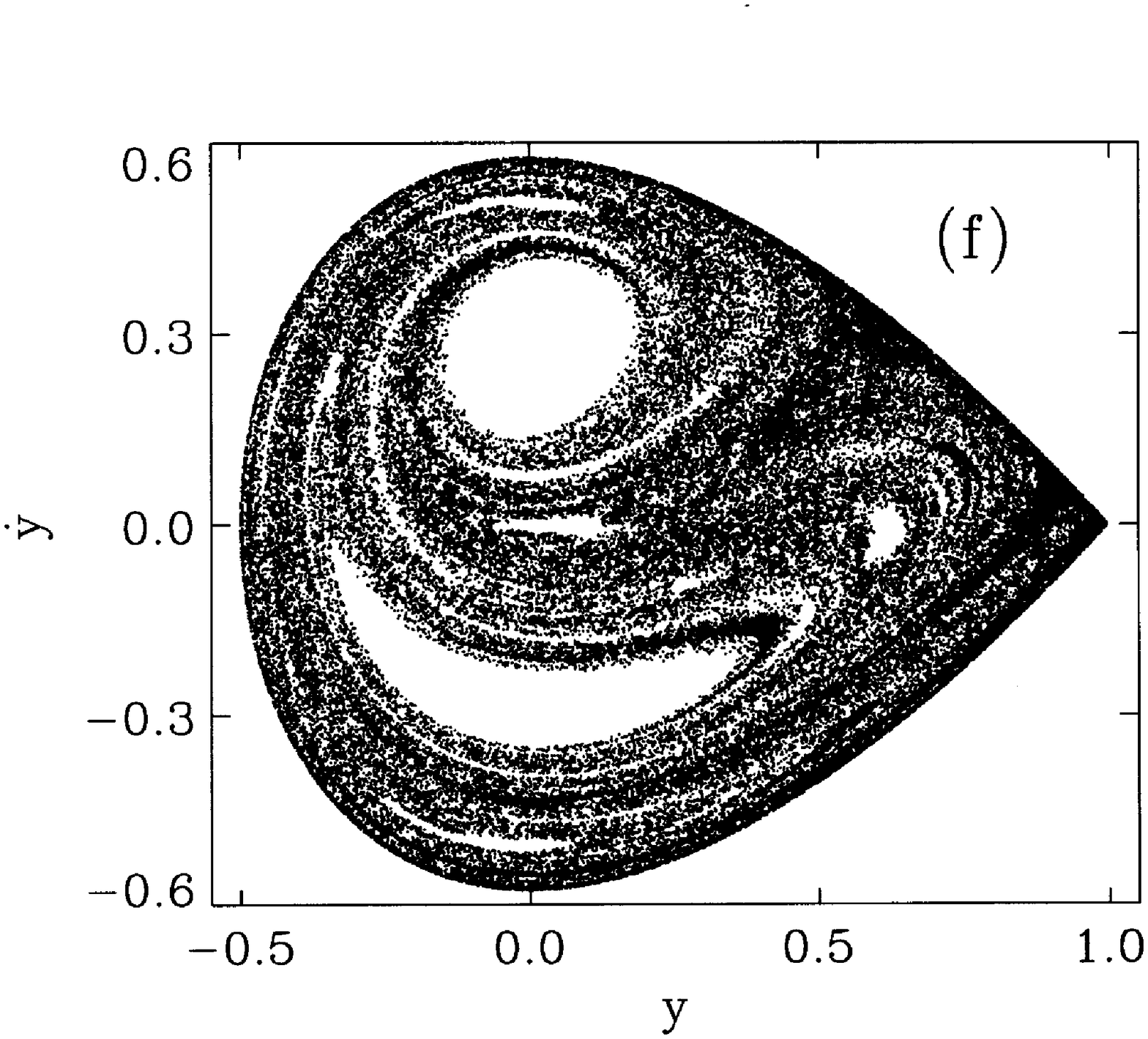}
           }
        \begin{minipage}{10cm}
        \end{minipage}
        \vskip -0.0in\hskip -0.0in

(f) The same for $t=25$.
\vspace{-0.0cm}
\end{figure}
\vfill\eject


\pagestyle{empty}
\begin{figure}[t]
\centering
\centerline{
        \epsfxsize=6 cm
        \epsffile{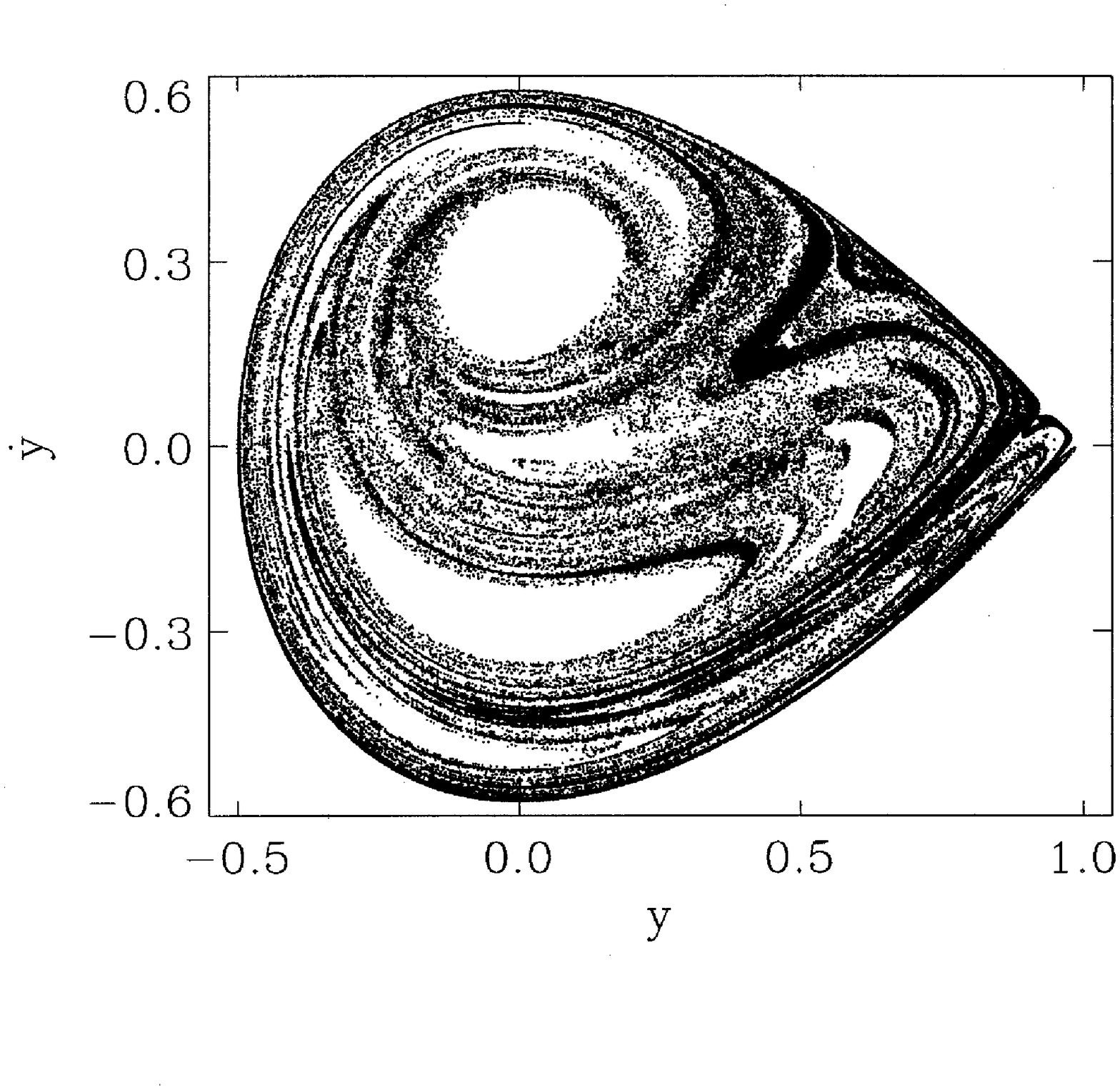}
           }
        \begin{minipage}{10cm}
        \end{minipage}
        \vskip -0.0in\hskip -0.0in

\caption{A surface of section analogous to Fig 5 c, generated at $t=10$ for
an ensemble of orbits with ${\epsilon}=1.06$.}
\vspace{-0.0cm}
\end{figure}


\pagestyle{empty}
\begin{figure}[t]
\centering
\centerline{
        \epsfxsize=8 cm
        \epsffile{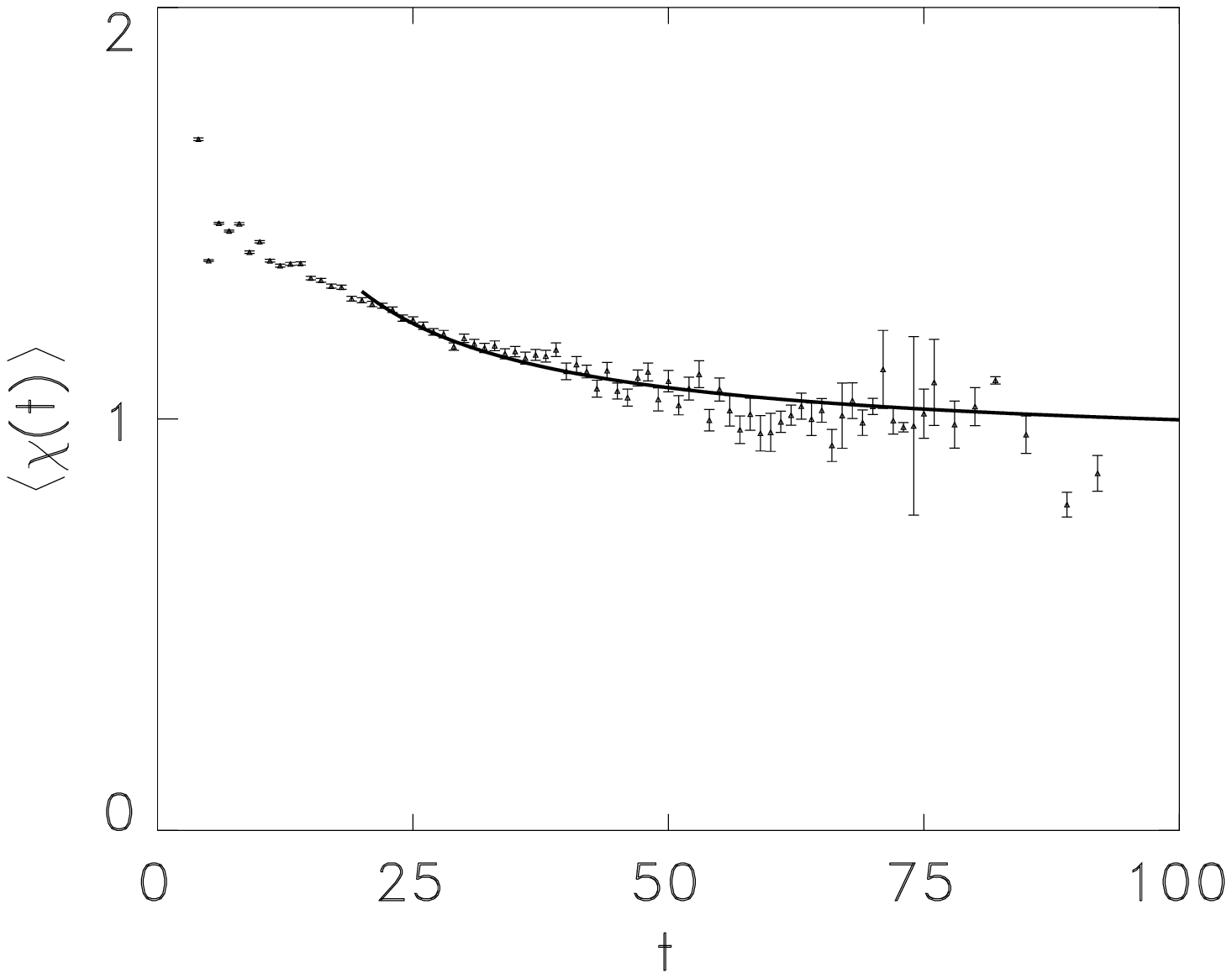}
           }
        \begin{minipage}{10cm}
        \end{minipage}
        \vskip -0.0in\hskip -0.0in

\caption{ (a) The escape probability, $P(t)$, computed for the same initial 
conditions as in Fig. 3 (c), now allowing for much longer times. The solid 
curve exhibits a power law fit, $P{\;}{\propto}{\;}t^{-\mu}$, with 
${\mu}=0.37.$ (b) The same data on a semilog plot, analysed as for Fig. 2b.}
\vspace{-0.0cm}
\end{figure}

\pagestyle{empty}
\begin{figure}[t]
\centering
\centerline{
        \epsfxsize=8 cm
        \epsffile{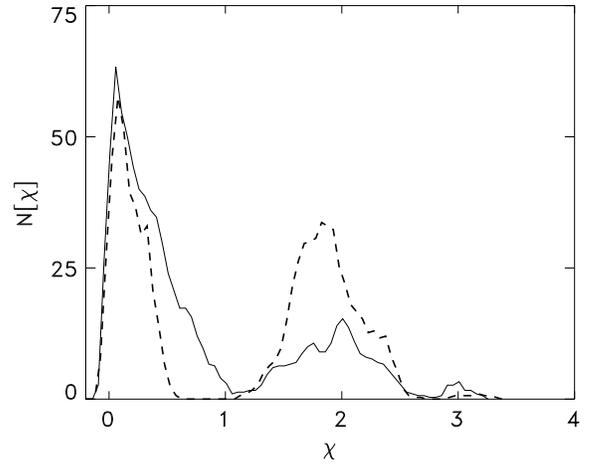}
           }
        \begin{minipage}{10cm}
        \end{minipage}
        \vskip -0.0in\hskip -0.0in

\caption{ Distributions of short time Lyapunov exponents for the intervals
$15<t<20$ (solid curve) and $45<t<50$ (dashed curve) for orbits that escape 
through the Lyapunov curves at consequent $t=50$. }
\vspace{-0.0cm}
\end{figure}\end{document}